\documentclass[%
 aps,
rsi,%
 amsmath,amssymb,
preprint,%
in-line,
longbibliography
]{revtex4-1}
\bibpunct{[}{]}{,}{n}{}{} 
\usepackage{graphicx}
\usepackage{dcolumn}
\usepackage{bm}
\usepackage[usenames]{color}

\begin{document}

\preprint{}

\title[]{Effect of multielectron polarization in strong-field ionization of the oriented CO molecule}

\author{Mahmoud Abu-samha}
 \email{Mahmoud.Abusamha@aum.edu.kw}
 \affiliation{College of Engineering and Technology, American University of the Middle East, Kuwait.}

\author{Lars Bojer Madsen}
\affiliation{Department of Physics and Astronomy, Aarhus University, 8000 Aarhus C, Denmark.}%

\date{\today}

\begin{abstract}
We discuss the mechanism by which the inclusion of multielectron polarization improves the solution of the time-dependent Schr\"{o}dinger equation (TDSE) for the oriented CO molecule in a strong external laser pulse within the single-active-electron (SAE) approximation. A challenging problem of using the SAE approximation is that the active electron, instead of undergoing ionization, may be  driven by the external field to lower-lying orbitals. For the oriented CO molecule, dipole coupling to the lower-lying bound states of the potential depends on the orientation angle, thereby affecting the orientation-dependent ionization dynamics. By including multielectron polarization, the external field is turned off within the molecular radius, thereby minimizing dipole coupling of the highest occupied molecular orbital to the lower-lying states of the potential. We discuss how turning off the external field within an appropriate molecular radius without accounting for the induced dipole term in the effective potential beyond this radial cut-off distance,  constitutes an effective and accurate approach to describe strong-field ionization of CO and to minimize dipole coupling to lower-lying bound states.

%
\end{abstract}

\pacs{}
\keywords{}
\maketitle

\section{Introduction}
For many-electron atoms and molecules in strong laser fields, a commonly used approximation for solving the time-dependent Schr{\"o}dinger equation (TDSE) is the single-active-electron (SAE) approximation in which the one-electron TDSE is solved for the least bound electron in the effective potential generated by the nuclei, the remaining electrons and the external field. Numerical methods are available in the literature for deriving time-independent SAE potentials for atoms~\cite{TongJPB05,PhysRevLett.81.1207} and molecules~\cite{PhysRevA.81.033416,PhysRevA.81.063406,doi:10.1080/00268970701871007,PhysRevLett.104.223001} based on quantum chemistry calculations. A challenging problem of using the SAE approximation is that the active electron may end up in one of the field-free lower-lying orbitals, and hence, the ionization dynamics will be  affected by dipole coupling to lower-lying bound states of the atomic or molecular potential. For atoms such as Ar~\cite{PhysRevLett.81.1207}, this problem has been circumvented by applying a hardcore boundary in the core region and performing the time propagation outside the core. For molecules in intense laser fields, a SAE treatment was carried out within the framework of time-dependent Hartree-Fock (TDHF) theory, in which the active orbital is constrained to remain orthogonal to the field-free lower-lying orbitals~\cite{PhysRevLett.104.223001}.  In TDHF, an extention to an all electron treatment provides a natural solution to this problem: if all electrons were driven by the field, then, due to the  unitary  evolution  under  the  time-dependent Hamiltonian, all orbitals remain orthogonal at all times without the need of external orthogonality constraints~\cite{PhysRevLett.111.163001}. 

\begin{figure}
 {\includegraphics[width=0.5\textwidth]{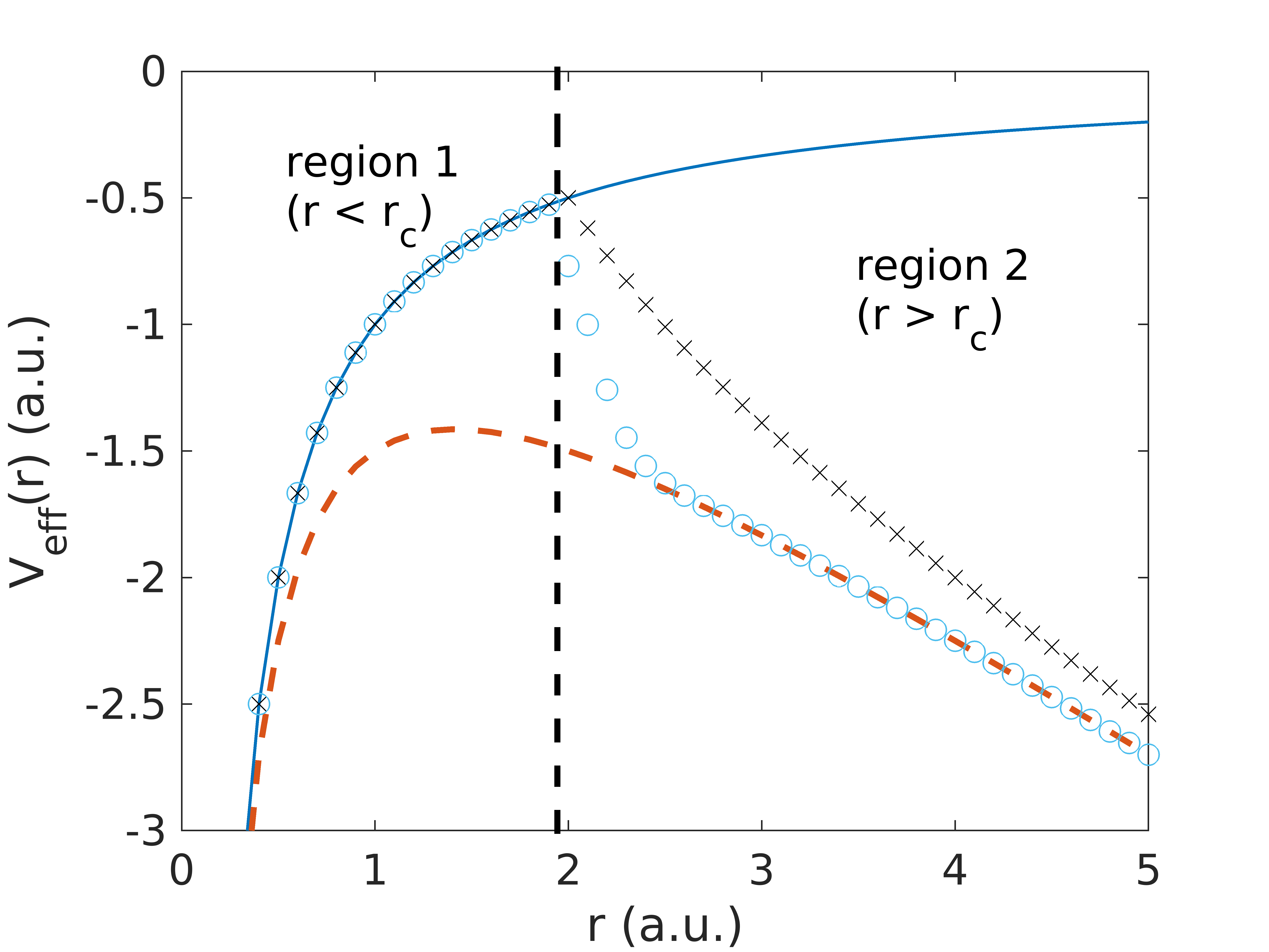}}
 \caption{Illustration of a field-free (solid line) and field-dressed potential. The dashed line (crosses) denote the field-dressed potential excluding (including) multielectron polarization, whereas the circles denote a field-dressed potential where the external field is turned off at $r < r_c= 2$~a.u. and the induced dipole term of the cation is neglected. To be able to see clearly the difference between the potentials on a linear scale, the illustration is for a system with a very large polarizability and a field strength of $0.5$ a.u.}
\label{Veff}
\end{figure}

In small multielectron atoms and molecules, the performance of the SAE model is widely accepted, and has recently been benchmarked against accurate calculations of ionization rates for rare gases (He, Ne, and Ar) and the H$_2$ molecule using a hybrid coupled channels approach including multielectron effects~\cite{Majety_2015b,Majety_2015a}. For atoms with large polarizabilities of their cations (Mg and Ca), effects of multielectron polarization (MEP) on total ionization yields and photoelectron momentum distributions were reported based on semiclassical calculations~\cite{PhysRevA.98.023406,PhysRevA.94.013415}. In Fig.~\ref{Veff}, we illustrate how MEP affects the potential felt by the ionized electron: The external field is turned off within the atomic/molecular radius (region 1 in Fig.~\ref{Veff}). At intermediate distances (region 2 in Fig.~\ref{Veff}), the effective potential is modified by an induced dipole term which depends on the polarizability of the cation. As can be seen from Fig.~\ref{Veff}, the effect of MEP in region 2 is to increase the tunneling barrier and shift the tunneling exit to a larger $r$ value, thereby reducing the total ionization yield as predicted by semiclassical theory~\cite{PhysRevA.91.033409,PhysRevA.98.023406}. The turning off of the laser-electron interaction at small distances is a consequence of the polarization of the remaining electrons. The electric field associated with this polarization counteracts the external field, and leads to a cancellation of laser-electron interaction. These points were discussed already in Ref.~\cite{doi:10.1080/09500340601043413}.


For molecules in strong laser fields, inclusion of MEP effects is essential for calculations of ionization rates~\cite{Holmegaard2010,PhysRevLett.111.163001,Majety_2015b,PhysRevA.96.053421} and high harmonic generation~\cite{Smirnova2009,Ferre2015,PhysRevA.76.031404}. Effects of MEP have recently been investigated for small molecules, including CO, NO and CO$_2$~\cite{Majety_2015b,PhysRevA.96.053421}. 
The interest in strong-field ionization of oriented targets, including the polar CO molecule, dates back to the past decade, and was motivated by advances in field-free orientation of molecular targets~\cite{Holmegaard2010,PhysRevLett.103.153002}. Already in 2010, angular distributions from polar molecules including CO were available from Stark-corrected molecular tunneling theory and TDSE calculations within the SAE approximation without MEP~\cite{PhysRevA.82.043413}. The TDSE calculations were performed for CO in a linearly polarized laser containing one cycle with a frequency corresponding to 800-nm wavelength and a laser intensity of $10^{14}$~W/cm$^2$. Both TDSE and Stark-corrected molecular tunneling theory predicted the maximum ionization yield when the laser field points from O to C. Likewise the weak-field asymptotic theory of tunneling ionization~\cite{WFAT} predicts a maximum in the rate when the field points from O to C~\cite{PhysRevA.85.053404}. 
Later on, experimental molecular-frame photoelectron momentum distributions (PMDs) of oriented CO became available and ionization of multiple orbitals was identified~\cite{PhysRevLett.108.183001}. In the experiment, the CO molecules were ionized by a 35-fs circularly polarized laser pulse with a central wavelength of 790 nm and intensity 4$\times10^{14}$~W/cm$^2$. The experimental results revealed a maximum ionization yield when the peak electric field points from C to O. It immediately became evident that the TDSE treatment within the SAE approximation needed revision. 



In Ref.~\cite{PhysRevLett.111.163001} strong-field ionization of the oriented CO molecule was investigated using all-electrons time-dependent Hartree-Fock theory (TDHF). According to Ref.~\cite{PhysRevLett.111.163001}, the orientation-dependent ionization of the CO molecule is strongly affected by MEP. In addition to the all-electron TDHF calculations, TDHF calculations within the single-active-orbital (SAO) approximation were also presented in Ref.~\cite{PhysRevLett.111.163001}. In the SAO model, only the HOMO electrons are propagated, while other orbitals remain frozen. This method, unlike the SAE approximation, requires exact calculations of direct and exchange potentials between the frozen core and the active HOMO electrons at each time step. Notice that the SAO approximation is similar to the SAE approximation in that an MEP term must be added to the effective potential in order for the SAO model to reproduce the experimental ionization yields~\cite{PhysRevLett.108.183001} for the CO molecule. This study has renewed the interest in the CO molecule as a target for strong-field ionization and high-harmonic generation, now with attention to the effect of MEP~\cite{PhysRevA.95.023407,0953-4075-51-3-034001,Song:17,PhysRevA.97.043405,Majety_2015b}.


A successful treatment of MEP within the SAE approximation was demonstrated recently in Ref.~\cite{PhysRevA.95.023407}, which promotes the SAE approximation as a more favorable alternative to the SAO method, because of its low computational costs. In Ref.~\cite{PhysRevA.95.023407}, the effect of MEP was investigated for polar (CO and NO) and nonpolar (N$_2$, O$_2$, and CO$_2$) molecules, based on TDSE treatment within the SAE approximation. For CO, for example, the multielectron effect is represented by an induced dipole term which includes the static polarizability of the CO$^+$ cation.
According to Ref.~\cite{PhysRevA.95.023407}, accounting for MEP was crucial for producing total ionization yields (TIYs) for CO molecule in qualitative agreement with the experiment. 


In this paper, we discuss in detail the improved performance of the TDSE method upon inclusion of MEP, with evidence from the case study of the CO molecule. As mentioned earlier, a major drawback of the SAE approximation is dipole coupling of the HOMO electron to the lower bound states of the potential, in particular the HOMO-2 orbital of the CO molecule as will be shown in Sec.~\ref{res}. The dipole matrix elements are strongly dependent on molecular orientation, and the total ionization yield at different orientation angles will be strongly affected by the dipole transitions and shifting of population from the HOMO to the lower-lying orbitals. Upon inclusion of MEP, this dipole coupling becomes negligible because the external field is turned off within the molecular radius (denoted by $r_c$ in Fig.~1). By extending our our TDSE method with MEP treatment following the approach of Refs.~\cite{PhysRevA.95.023407,0953-4075-51-10-105601}, we will show that the long-range correction term (region 2 in Fig.~\ref{Veff}) is not needed to produce the correct trend in TIYs for the CO molecule, as long as the external field is simply turned off within the molecular radius. 


The computational details are summarized in Sec.~\ref{compdet}, followed by results and discussion in Sec.~\ref{res} and conclusions in Sec.~\ref{conc}. Atomic units (a.u.) are used throughout unless otherwise stated.

\section{Theoretical and computational models}
\label{compdet}

\subsection{Extending the TDSE method with MEP}

For polar molecules in the case where MEP is neglected, the potential describing the interaction of the SAE with the frozen core and the time-dependent external field is given asymptotically as~\cite{PhysRevA.82.043413}
\begin{equation}
\label{saepot1}
V(\vec{r},t) = \vec{r}\cdot\vec{E}(t) -\frac{Z}{r} - \frac{\vec{\mu}_p\cdot \vec{r}}{r^3} -   \cdots
\end{equation}
where $Z=1$ is the charge of the cation, and $\vec{\mu}_p$ is the permanent dipole of the cation. The effect of MEP on strong field ionization was developed in Refs.~\cite{PhysRevLett.95.073001,doi:10.1080/09500340601043413,PhysRevA.82.053404,Holmegaard2010}. Accordingly, the effective potential felt by the tunneled electron in an external field may be defined asymptotically as~\cite{PhysRevLett.95.073001,PhysRevA.82.053404}
\begin{equation}
\label{saepot2}
V_\text{eff}(\vec{r},t) = \vec{r}\cdot\vec{E}(t) -\frac{Z}{r}  - \frac{(\vec{\mu}_p + \vec{\mu}_{ind})\cdot \vec{r} }{r^3}
\end{equation}
where $\vec{\mu}_p$ and  $\vec{\mu}_{ind}$ are the permanent and induced dipoles of the cation. By comparing Eqs. (\ref{saepot1}) and (\ref{saepot2}), one can see that the SAE potential used in most TDSE calculations [Eq. (\ref{saepot1})] is missing the interaction term due to the induced dipole; that is ($- \vec{\mu}_{ind}\cdot \vec{r} /r^3$) is missing. To include MEP in our TDSE method, we adopted the approach layed out in Ref.~\cite{0953-4075-51-10-105601}. The MEP term is defined as $\left(- \vec{\mu}_{ind}\cdot \vec{r} /r^3 = -\alpha_{||} \vec{E}(t)\cdot\vec{r}/r^3 \right)$ where $\alpha_{||}$ is the static polarizability of the ion parallel to the laser polarization axis. A cutoff radius is chosen close to the core at a radial distance
\begin{equation}
   r_{c}=\alpha_{||}^{1/3}
   \label{r_c}
\end{equation}
such that the MEP cancels the external field at $r\le r_c$~\cite{0953-4075-51-10-105601,PhysRevLett.95.073001,doi:10.1080/09500340601043413}. As mentioned in Sec.~I, turning off the laser-electron interaction at small distances ($r\le r_c$) is a consequence of the polarization of the remaining electrons: The electric field associated with this polarization counteracts the external field and leads to a cancellation of the laser-electron interaction~\cite{doi:10.1080/09500340601043413}.

Implementing the MEP term in the TDSE method is straightforward in the length gauge (LG). In our approach, without MEP, the interaction term is expressed in the LG at each radial grid point $r_i$ as~\cite{Kjeldsen_phd} 
\begin{equation}
\label{V_LG}
    V_{LG}^{Ext}(r_i,t)=E(t) r_i \cos(\theta) = E(t) \sqrt{\frac{2}{3}}r_i \bar{P}_{1}(\zeta).
\end{equation}
where $E(t)$ is the electric field at time $t$, $\bar{P}_1$ is a normalized Legendre function, and $\zeta=\cos(\theta)$ where $\theta$ is the polar angle of the electron coordinate $\vec{r}$. Upon inclusion of the MEP term, the interaction term then reads
\begin{equation}
\label{Eq2}
    V_{LG}^{Ext}(r_i,t)= \begin{cases}
  \left(1-\frac{\alpha_{||}}{r_i^3}\right)E(t)\sqrt{\frac{2}{3}}r_i \bar{P}_{1}(\zeta);~r > r_c \\ 
  0;~r \le r_c \\
  \end{cases}
  \end{equation}
where the interaction term is zero at $r\le r_c$ because the external field is counteracted by the MEP~\cite{PhysRevLett.95.073001,doi:10.1080/09500340601043413}.

The electric field $E(t)$, linearly-polarized along the lab-frame $z$-axis, is defined as
\begin{equation}
    \vec{E}(t) = -\partial_t A(t) \hat{z}= -\partial_t \left(\frac{E_0}{\omega}\sin^2(\pi t/\tau)\cos(\omega t+\phi) \right)\hat{z},
    \label{E_field}
\end{equation}
where $E_0$ is the field amplitude, $\omega$ the frequency, and $\phi$ the carrier-envelope phase (CEP) for a laser pulse with duration $\tau$. The TDSE calculations were performed at laser frequency $\omega=0.057$~au corresponding to 800~nm wavelength, CEP value $\phi=-\pi/2$, and pulse durations and intensities as specified below. 

To include MEP in the TDSE calculations for CO, the polarizability [$\alpha_{||}$ in Eq.~(\ref{Eq2})] was determined for the CO$^+$ cation at the frozen geometry of the neutral CO molecule (with a C--O bond distance of 2.13~a.u.). The polarizability was determined from quantum chemistry calculations within the framework of density functional theory employing the local spin-density approximation (LSDA)~\cite{doi:10.1139/p80-159} and Aug-cc-pVDZ basis set~\cite{doi:10.1063/1.456153}. The CO$^+$ cation has a static polarizability tensor with non-zero components $\alpha_{xx}=\alpha_{yy}=7.88$~a.u. and $\alpha_{zz}=12.39$~a.u. The dynamic polarizability was computed for the CO$^+$ cation at an external field frequency ($\omega$=0.057~a.u.) corresponding to 800-nm wavelength, and the produced polarizability tensor has the following non-zero components $\alpha_{xx}=\alpha_{yy}=8.22$~a.u. and $\alpha_{zz}=12.63$~a.u. We notice that the static and dynamic polarizability of the CO$^+$ cation are very similar, the difference is less than 5\%.

In the TDSE calculations for the CO molecule at orientation angle $\beta$, the polarizability along the laser polarization axis ($\alpha_\parallel$)  may be computed as $\alpha_\parallel=\alpha_{zz} \cos^2(\beta) + \alpha_{xx} \sin^2(\beta)$. At $\beta=0^\circ$, the molecular and lab-frame $z$-axes are aligned, and in this case $\alpha_\parallel=\alpha_{zz}=12.39$~a.u. At $\beta=90^\circ$, by contrast, the molecular frame $x$-axis is aligned with and lab-frame $z$-axis, and in this case $\alpha_\parallel=\alpha_{xx}=7.88$~a.u.

\subsection{Turning off the external field within the molecular radius}
\label{sec2}
In order to turn off the external field at $r < r_c$ (region 1 of Fig.~\ref{Veff}) while neglecting the effect of the MEP term at $r > r_c$ , the external field was defined in the LG as
\begin{equation}
V_{LG}^{Ext}(r_i,t)= \begin{cases}
  \Omega(r)E(t)\sqrt{\frac{2}{3}}r_i \bar{P}_{1}(\zeta);~r > r_c \\ 
  0;~r \le r_c \\
  \end{cases}
\label{V_turn_off}
\end{equation}
where $\Omega(r)$ is a radial scaling function  defined as
\begin{equation}
\label{eq:omega}
    \Omega(r)=\left(\frac{1}{2}+\frac{1}{2}\tanh{\left[\frac{r-r_c-d}{s}\right]}\right),
\end{equation}
and the factors $s$ and $d$ determine the steepness of the turnoff of the external field at the cutoff radius $r_c$. The implementation of this cutoff potential is demonstrated for the model potential shown in Fig.~\ref{saepot1}. In the TDSE calculations for CO, we set $s=1.0$ and $d=1.0$.

\subsection{SAE potential and HOMO of CO}

The SAE potential describing CO was determined from quantum chemistry calculations following the procedure in Ref.~\cite{PhysRevA.81.033416}. The molecule is placed along the molecular-frame $z$-axis such that the center-of-mass coincides with the origin and the O atom points in the positive $z$ direction. The SAE potential of CO was expanded in partial waves as $V(\mathbf{r})=\sum_{l,m=0}^{l_{max}}V_{l0}(r)Y_{l0}(\theta,\phi)$ where $m=0$ since the molecule is linear. The expansion was truncated at $l_{max}=20$. Based on our SAE potential for CO, the highest occupied molecular orbital (HOMO) of CO is the $5\sigma$ with energy -0.542~au in reasonable agreement with the literature value of -0.555~au~\cite{KOBUS19937}.

We follow the split-operator spectral method of Hermann and Fleck~\cite{PhysRevA.38.6000}(see Ref.~\cite{Kjeldsen_phd} for details of our implementation) to obtain the wavefunction of the HOMO of the CO molecule. Starting with a guess initial wavefunction, denoted $\Psi(r,t=0)$ and the SAE potential for the CO molecule, we perform field-free propagation for 1000 a.u. and save the wavepacket every 1.0 a.u. From the time-dependent wavefunction, $\Psi(r,t)$, a bound-state spectrum $\mathcal{P}(E)$ is produced with the aid of the auto-correlation function $\mathcal{P}(t)$, as
\begin{equation}
\mathcal{P}(E) =     \frac{1}{T}\int_{0}^{T}dt~\omega(t)~\exp(iEt)~\mathcal{P}(t)
\end{equation}
where $w(t)$ is the Hanning window function~\cite{PhysRevA.38.6000} and where $\mathcal{P}(t)= \langle\Psi(r,t=0)|\Psi(r,t) \rangle$. Once the orbital energy of the HOMO, $E_\text{HOMO}$, is well resolved in the $\mathcal{P}(E)$ spectrum, the corresponding wavefunction can be constructed and normalized as follows
\begin{equation}
\label{wfc_homo}
\Psi_\text{HOMO}(r) =     \frac{1}{T}\int_{0}^{T}dt~w(t)~\exp(iE_\text{HOMO}t )\Psi(r,t).
\end{equation}

In the TDSE calculations, the radial grid contains 4096 points and extends to 400~au. The size of the angular basis set is limited by $l_{max}=40$ (50 for convergence tests). The calculations were performed at orientation angles in the range $\beta=0-180$  with a step of 30$^\circ$. The PMDs were produced by projecting the wavepacket at the end of the laser pulse on Coulomb scattering states in the asymptotic region ($r>20$~a.u.), an approach that was validated in Ref.~\cite{Madsen2007} and recently applied for the molecular hydrogen ion~\cite{PhysRevA.94.023414,1742-6596-869-1-012009}.

\section{Results and Discussion}
\label{res}

\subsection{Extending the TDSE method with MEP within the SAE approximation}

Here, we study strong-field ionization of the polar CO molecule as a function of orientation angle $\beta$ and explore the effect of MEP on the orientation-dependence of TIYs within the SAE approximation. The CO molecule fixed at an orientation angle $\beta$ in the range $0^\circ-180^\circ$ was probed by a linearly polarized laser pulse containing 2 optical cycles, with a frequency ($\omega=0.057$~au) corresponding to 800-nm wavelength (at 800 nm, the duration of a single cycle is 110.2 a.u.), a laser peak intensity of 8.8$\times10^{13}$~W/cm$^{2}$ (one atomic unit of intensity corresponds to $3.51 \times 10^{16}$ W/cm$^2$), and a CEP value of $\phi=-\pi/2$. At $\beta=0^\circ$, the peak electric field points from C to O, whereas at $\beta=180^\circ$ the peak electric field points from O to C.  At the laser parameters considered here, the contribution to the TIYs from the HOMO-1 orbital is insignificant at all orientations, in agreement with previous theoretical calculations~\cite{PhysRevLett.111.163001}, and will be omitted in the present study.

\begin{figure*}
 {\includegraphics[width=1.0\textwidth]{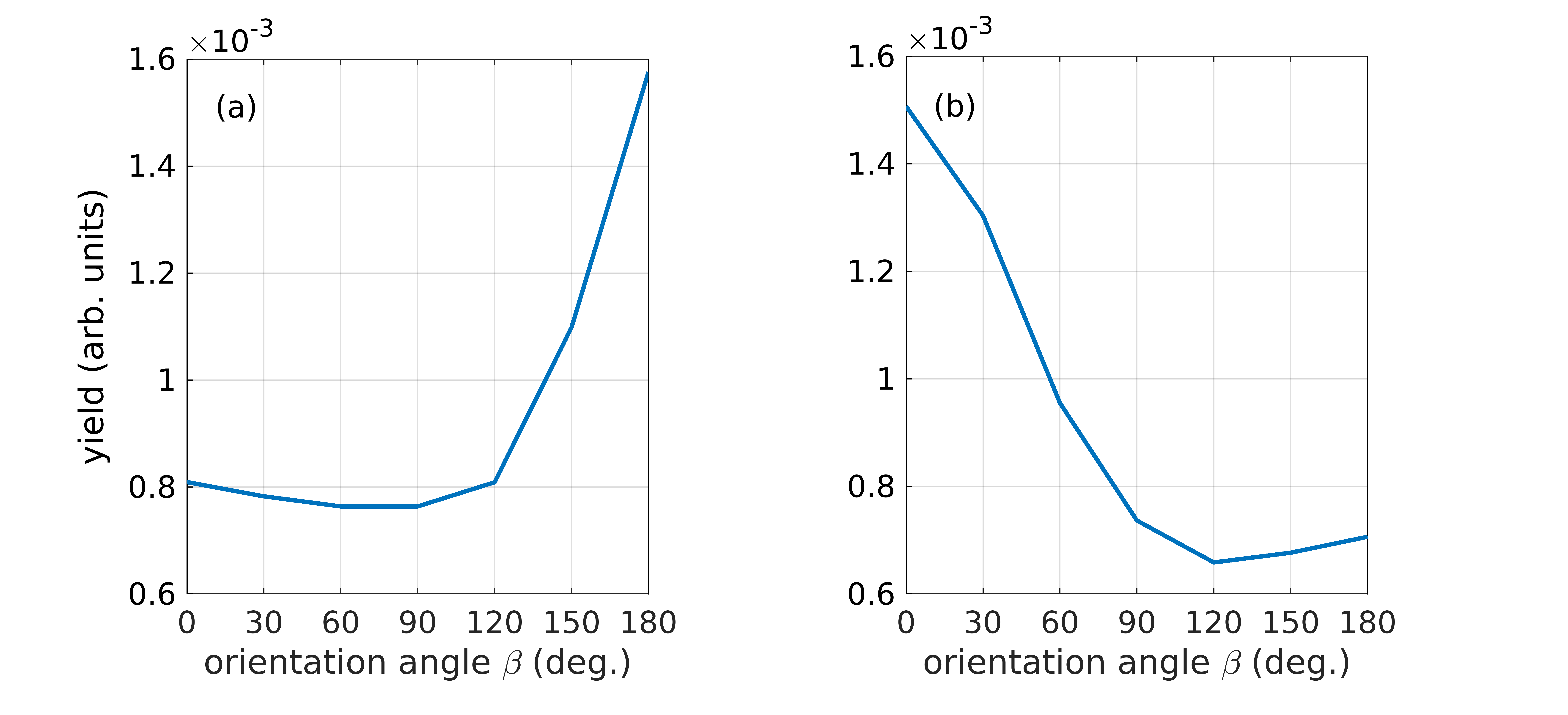}}
 \caption{Orientation-dependent TIYs from the $5\sigma$ HOMO of the CO molecule, based on TDSE calculations (a) without the multielectron polarization term corresponding to the external potential in Eq.~\eqref{V_LG} and (b) with the multielectron polarization term corresponding to the external potential in Eq.~\eqref{Eq2}. The CO molecule was ionized by a 2-cycle linearly-polarized laser pulse with a wavelength of 800~nm ($\omega=0.057$ a.u.) and an intensity of 8.8$\times10^{13}$~W/cm$^2$ (0.0025 a.u.). At $\beta = 0^\circ$ the peak field of the pulse points from C to O.}
\label{co_tiy}
\end{figure*} 

In Fig.~\ref{co_tiy} we plot the TIYs from the HOMO of the CO molecule at different orientation angles $\beta$. The results in Fig.~\ref{co_tiy}(a) are produced from TDSE calculations without accounting for MEP and describing the external field by Eq.~\eqref{V_LG}. In this case, the TDSE results predict a maximum TIY at $\beta=180^\circ$, that is, when the maximum of the electric field points from O to C, and a minimum TIY at $\beta=90^\circ$. A comparison of these results to the experimental measurements of Ref.~\cite{PhysRevLett.108.183001}, where a maximum yield is obtained at $\beta=0^\circ$ and a minimum yield at $\beta=120^\circ$, shows that extending the TDSE method beyond the single-active-electron approximation is essential. Let us now turn the discussion to the results in Fig.~\ref{co_tiy}(b), where multielectron effects are accounted for in the TDSE calculations as described in Sec.~II A [see Eq.~\eqref{Eq2}]. In this case, the TDSE results predict a maximum TIY when the maximum of the electric field points from C to O ($\beta=0^\circ$) and a minimum at $\beta=120^\circ$. The TDSE results are now in good agreement with the experimental measurements~\cite{PhysRevLett.108.183001} and recent theoretical calculations~\cite{PhysRevA.95.023407}.

\begin{figure}
 {\includegraphics[width=0.5\textwidth]{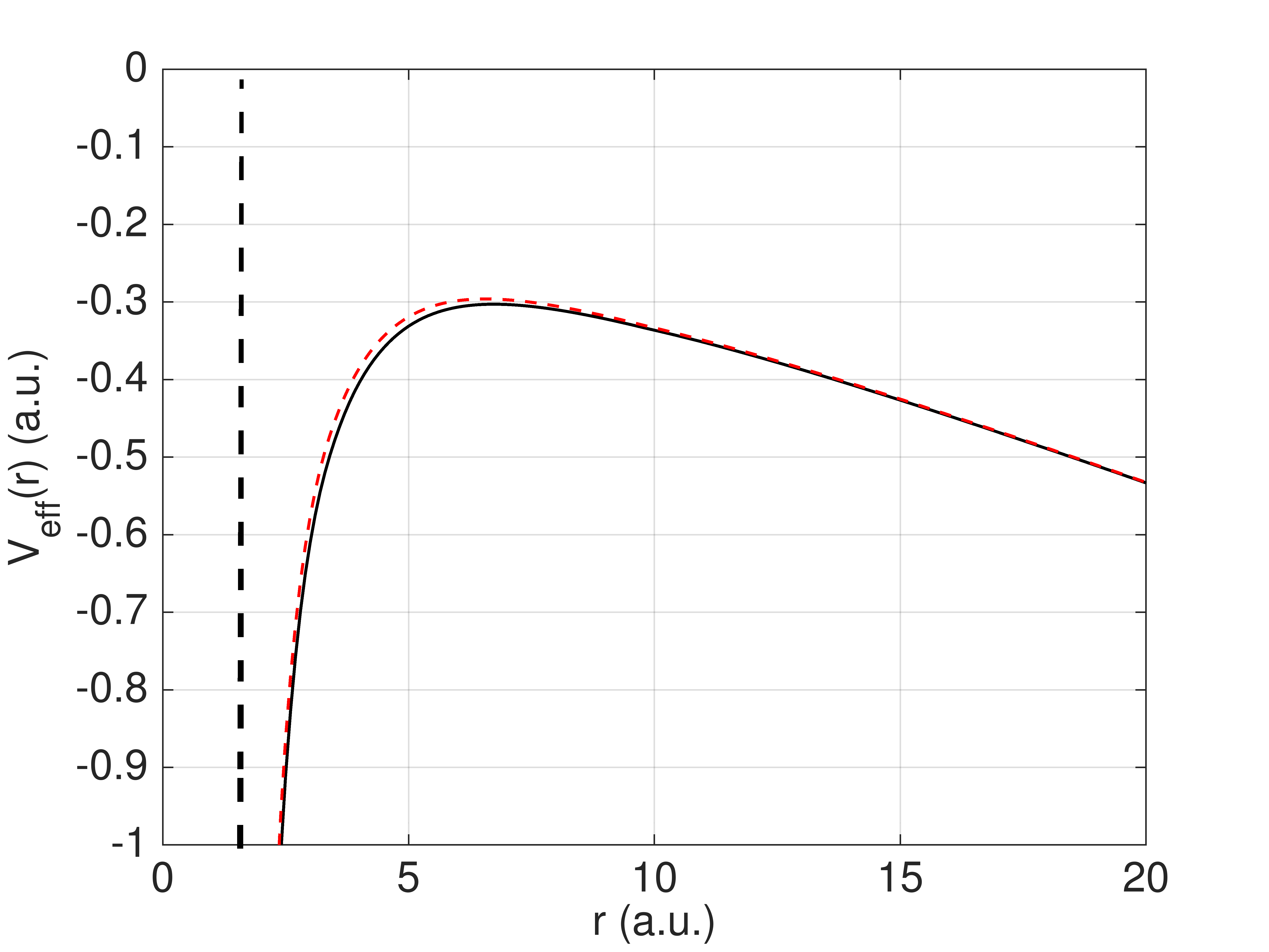}}
 \caption{Field-dressed potential with (dashed line) and without (solid line) account for MEP. The field-dressed potentials were generated at an electric field value of -0.05~a.u. corresponding to a laser peak intensity of 8.8$\times10^{13}$~W/cm$^2$, and CEP $\phi=\pi/2$ [see Eq.~\eqref{E_field}], and the external field is turned off within $r_c = 2.1$~a.u., which is indicated by the vertical dashed line.}
\label{CO_Veff}
\end{figure}

\subsection{Turning off the external field at short radial distances}

From the preceding discussion, the performance of the TDSE method, within the SAE approximation, is substantially improved by including an MEP term in the effective potential [see Eq.~(\ref{Eq2})] while keeping a low computational cost. Let us now discuss how the effective potential for the CO molecule is improved by including MEP. In Fig.~\ref{CO_Veff}, we plot the field-dressed potential for the CO molecule, with and without the MEP effect. As can be seen in Fig.~\ref{CO_Veff}, the effect of MEP at $r > r_c$ is not significant ($r_c$ of Eq.~\eqref{r_c} is represented by a vertical dashed line in Fig.~\ref{CO_Veff}). Based on classical calculations of the tunneling exit following the approach of Ref.~\cite{PhysRevA.91.033409}, the tunneling exit changes from -11.33 a.u. to -11.67 a.u.  (10.8 a.u. to 10.9 a.u.) upon inclusion of MEP at an external field value of 0.05 a.u. (-0.05 a.u.). Evidently, the change of the tunnel exit due to inclusion of MEP cannot explain the difference in total ionization yields at the orientation angles $\beta=0^\circ$ and 180$^\circ$. In fact, as we will show below, the long-range part of the MEP (at $r > r_c$) can be neglected in the case of the CO molecule without much effect on the TIYs and their orientation dependence. The exact value of this induced dipole term of the potential is only important for systems with very large polarizabilities, as has been shown in strong-field ionization studies for both atoms (calcium in Ref.~\cite{PhysRevA.98.023406}) and molecules (naphthalene in Refs.~\cite{PhysRevLett.106.073001,Dimitrovski_2015}). For systems with smaller polarizabilities it is only necessary to account for the turn-off effect of the field at small radial distances.

\begin{figure*}
 {\includegraphics[width=1.0\textwidth]{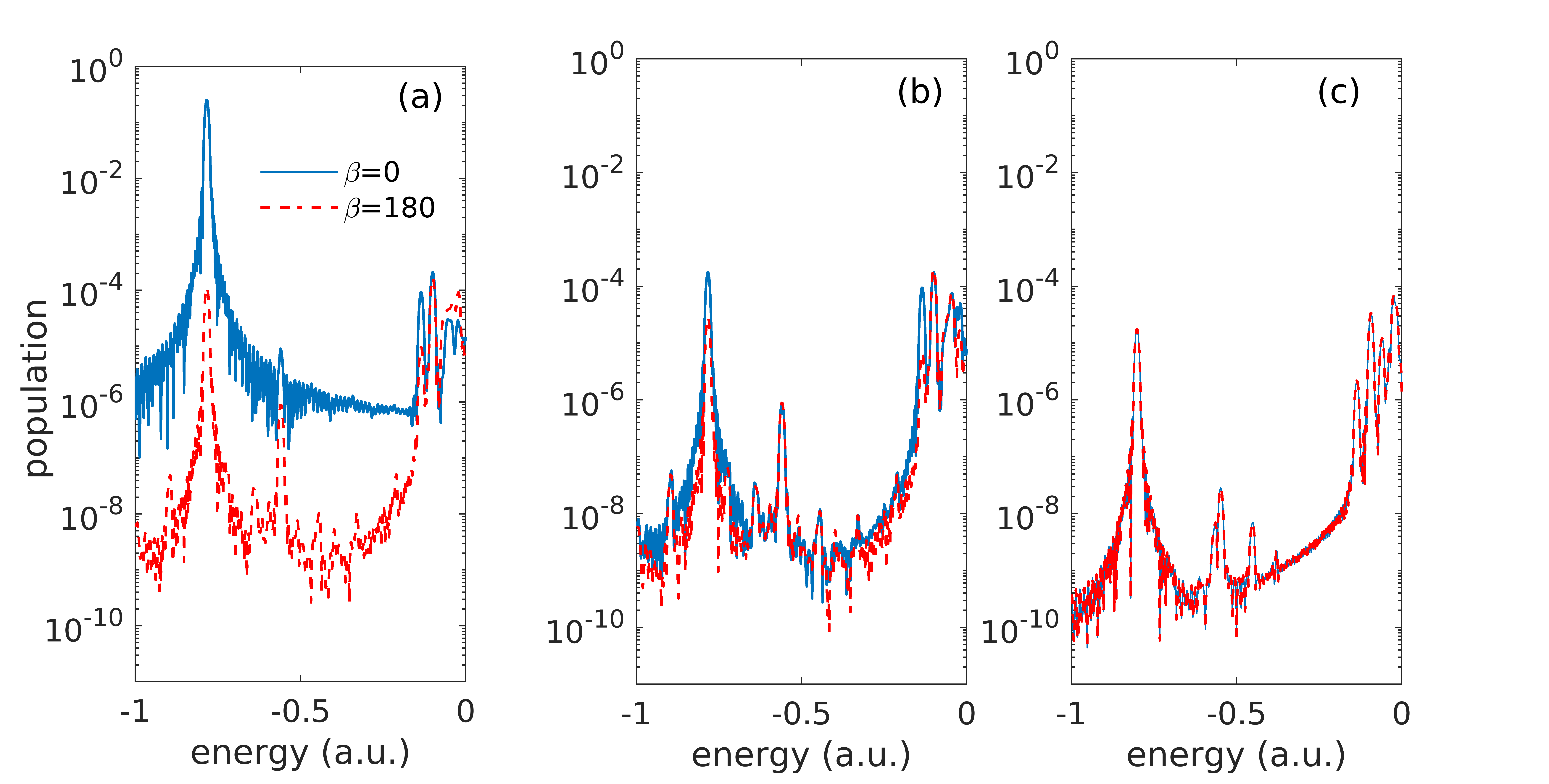}}
 \caption{Population (logarithmic scale) of bound states of the CO molecule at the end of a 2-cycle linearly-polarized pulse (a) without MEP effect [Eq.~\eqref{V_LG}], (b) with full MEP effect [Eq.~\eqref{Eq2}], and (c) when the external field is turned off within the molecular radius [Eq.~\eqref{V_turn_off}]. The laser pulse has a frequency (0.057~a.u.) corresponding to a wavelength of 800~nm and an intensity of 8.8$\times10^{13}$~W/cm$^2$. The HOMO was projected out from the wavepacket before producing the bound-state spectra. The full (blue) curve shows  the population of bound states computed at orientation angle $\beta=0$ whereas the dashed (red) curve shows their population at orientation angle $\beta=180$. The peak at $\sim -0.78$ corresponds to the H0MO-2, the peak at $\sim -0.55$ corresponds to HOMO, and the population at energies above -0.2 correspond to excited states. At $\beta = 0^\circ$ the peak field of the pulse points from C to O.}
\label{co_ati}
\end{figure*} 

Let us now pay attention to the vertical dashed line in Fig.~\ref{CO_Veff}, which indicates the radial cutoff distance below which the external field should be turned off. The cutoff radius $r_c=2.1$~a.u. was estimated based on the isotropic polarizability of the CO$^+$ cation ($\alpha$=9.38~a.u.), using Eq.~\eqref{r_c}~\cite{doi:10.1080/09500340601043413}. Our hypothesis is that the improved performance of the TDSE method for CO is caused simply by the switching off of the external field at radial distances within the molecular radius, that is at $r < r_c$ and not by the modification of the effective potential at $r > r_c$ by the induced dipole term of the potential. By turning off the external field within the molecular radius, we minimize the dipole coupling of the HOMO to the lower-lying bound states of the field-free potential, which in turn increases the accuracy of the TDSE results. This is supported by evidence of strong dipole coupling to lower-bound states of the CO molecule at orientation angles $\beta=0^\circ$ 
when the TDSE calculations are performed without MEP, as can be seen in Fig.~\ref{co_ati}(a). The population of bound states was determined by taking the Fourier transform of the auto-correlation function obtained from field-free propagation of the wavepacket, after the laser pulse has ended, analogously to the procedure described in Sec.~II C for generating the initial HOMO. The initial HOMO of CO was projected out from the wavepacket prior to field-free propagation. From Fig.~\ref{co_ati}(a), a significant population is found in the HOMO-2 orbital at an energy of -0.78~a.u. at orientation angle $\beta=0^\circ$. At $\beta=180^\circ$, by contrast, very little population is found in the HOMO-2.  In Fig.~\ref{co_ati}(b), the population of bound states was computed for the CO molecule, based on TDSE calculations including full MEP effect as described by Eq.~\eqref{Eq2}, where the external field is turned off at $r < r_c$ and an induced dipole term is applied at $r > r_c$. In this case, we find negligible population in the lower bound states (in particular the HOMO-2), compared to TDSE calculations without the MEP effect (where the external field is not turned off at short radial distances), cf. Fig.~\ref{co_ati}(a). Evidently, the dipole coupling to the lower bound states in the CO molecule is very significant and is orientation dependent. This, indeed, explains the failure of the TDSE method to reproduce the TIYs for the CO molecule within the SAE approximation without including the MEP term.

\begin{figure}
 {\includegraphics[width=0.5\textwidth]{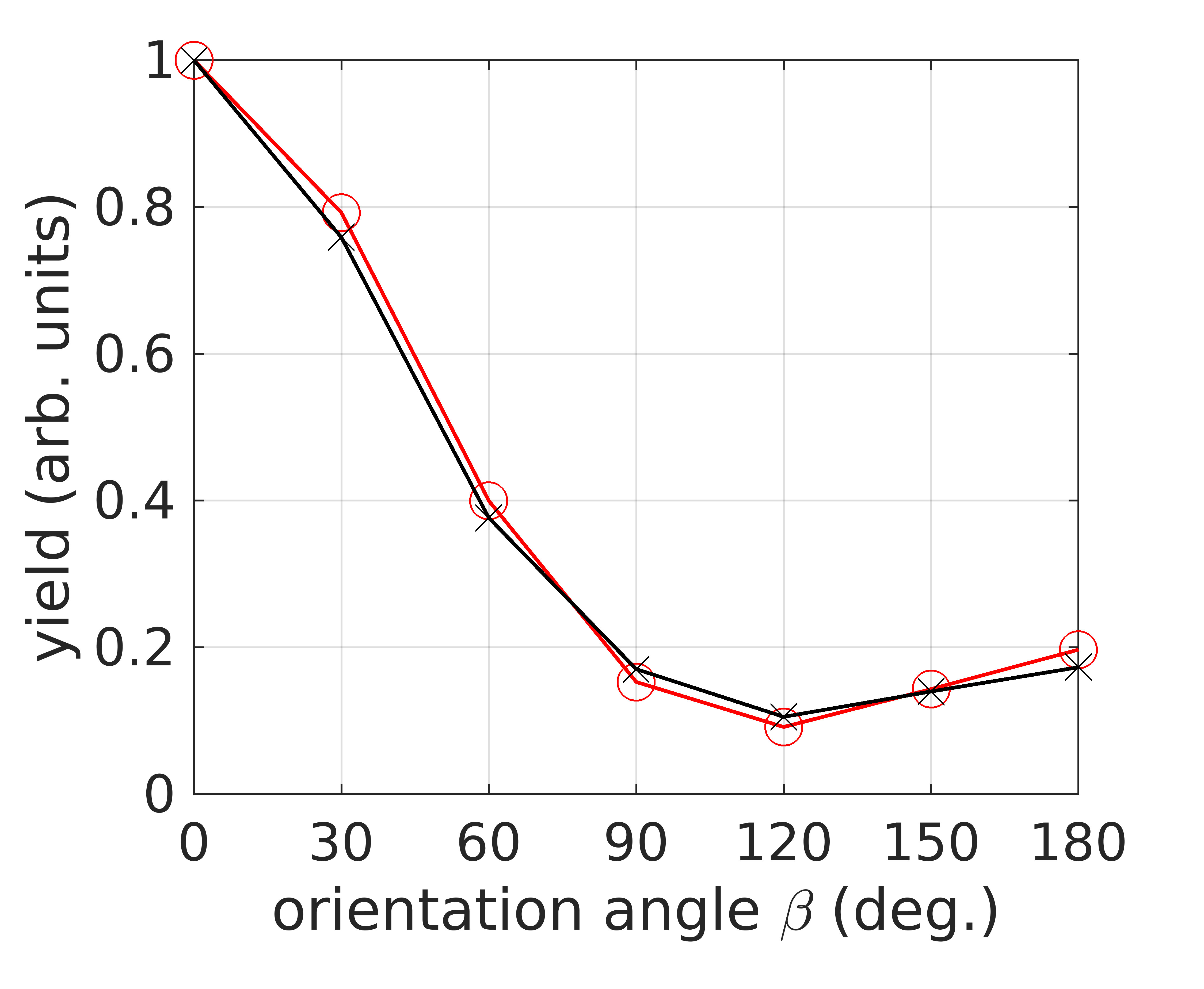}}
 \caption{TIYs from the HOMO of the oriented CO molecule from TDSE method within the SAE approximation. The full MEP effect [Eq.~\eqref{Eq2}] is represented by crosses, whereas the circles represent the case where the external field is turned off at $r < r_c$ [Eq.~\eqref{V_turn_off}]. In both cases, the CO molecule was probed by a 2-cycle linearly-polarized pulse with a frequency (0.057~a.u.) corresponding to a wavelength of 800~nm and an intensity of 8.8$\times10^{13}$~W/cm$^2$. At $\beta = 0^\circ$ the peak field of the pulse points from C to O.}
\label{co_tiy_tanh}
\end{figure}

 To provide further support to the preceding results, we computed the distribution of TIYs for the CO molecule with the external field turned off at $r < r_c$ while neglecting the MEP effect at $r > r_c$  (i.e. neglecting the induced dipole term). The external field is turned on smoothly around $r_c$ by applying a scaling function [see Eq.~(\ref{eq:omega})] described in Sec.~\ref{sec2}. The produced TIYs are compared to TDSE calculations with full MEP effect in Fig.~\ref{co_tiy_tanh}. In Fig.~\ref{co_tiy_tanh} the TIYs at orientation angle $\beta=0^\circ$ were normalized to unity. We note in passing that the new approach produces somewhat larger TIYs compared to TDSE calculations with full MEP effect, in agreement with previous predictions by the semiclassical theory~\cite{PhysRevA.91.033409,PhysRevA.98.023406}.  We see from the figure that the shapes of the TIYs are very similar as a function of $\beta$, in particular the proposed approach predicts accurately the orientation angles of maximum and minimum ionization yields. The population of bound states of the CO molecule calculated when the external field is turned off within the molecular radius  at $r < r_c$ and no induced dipole term is included beyond $r_c$ is shown in Fig.~\ref{co_ati}(c). With this approach, we produce quantitatively similar results compared to the TDSE calculations with full MEP [see Fig.~\ref{co_ati}(b)]:  we minimize the coupling of the HOMO to the lower lying-bound states of the CO molecule, meanwhile having minimal effect on the population in the excited states.  

\begin{figure*}
 {\includegraphics[width=1.0\textwidth]{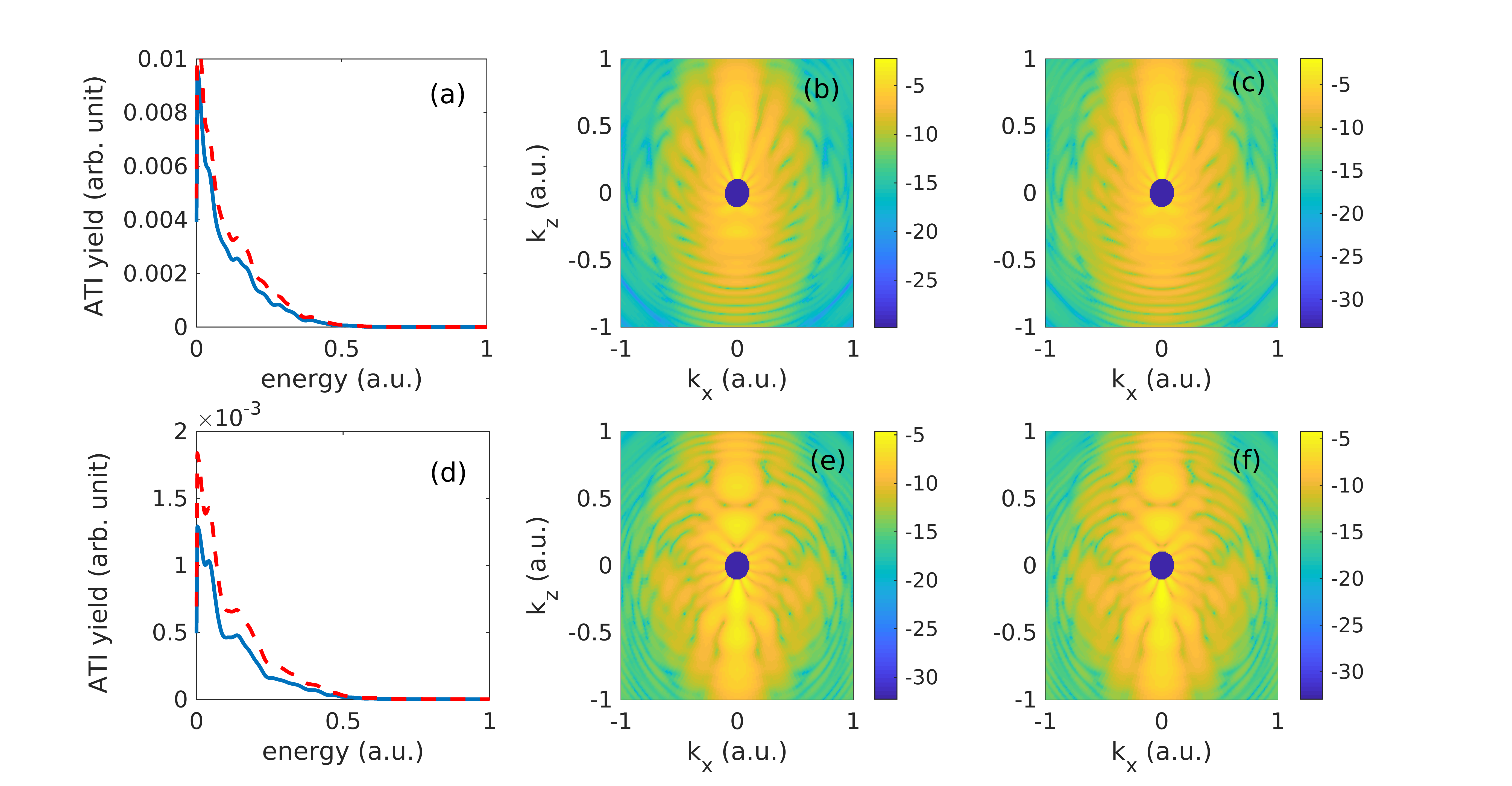}}
 \caption{ATI spectra and PMDs (the scale is logarithmic) from the HOMO of the oriented CO molecule at (a)-(c)$\beta=0^\circ$ and (d)-(f)$\beta=180^\circ$ from the TDSE within the SAE approximation. The CO molecule was probed by a 2-cycle linearly-polarized pulse with a frequency (0.057~a.u.) corresponding to a wavelength of 800~nm and an intensity of 8.8$\times10^{13}$~W/cm$^2$. In the ATI spectra, the solid curves represent full MEP effect, while the dashed curves represent the case where the external field is turned off at $r < r_c$ and the induced dipole potential is neglected at $r>r_c$. The PMDs in (b) and (e) account for full MEP effect whereas the PMDs in (c) and (f) correspond to the case where the external field is turned off at $r < r_c$ and the induced dipole potential is neglected at $r>r_c$.}
\label{co_pmds}
\end{figure*}

To address the effect at $r > r_c$ caused by neglecting the induced dipole term  caused by the response of the CO$^+$ cation to the external field on differential ionization observables, the above-threshold ionization (ATI) spectra and photoelectron momentum distributions (PMDs) were computed for the CO molecule at orientation angles $\beta=0^\circ$ and 180$^\circ$, and the results are compared to TDSE calculations with full MEP effect in Fig.~\ref{co_pmds}. From the ATI spectra in Figs.~\ref{co_pmds}(a) and (d), neglecting the induced dipole term results in a somewhat larger TIYs. Turning to the PMDs, neglecting the induced dipole term has no observable effect on the main features.

\section{Conclusions and outlook}
\label{conc}
By considering the case of the CO molecule, we have shown that a single-active-electron TDSE approach accounting for multielectron polarization effects as described by the potential in Eq.~\eqref{Eq2} captures the correct behavior of the strong-field induced total ionization yields with molecular orientation with respect to the dominant half cycle of the ionizing pulse. We have then considered a simpler potential, Eq.~\eqref{V_turn_off}, where the external field is simply turned off at radial distances smaller than $r_c$ of Eq.~\eqref{r_c}, below which the polarization of the electronic charge cancels the external field, and the effect of the induced dipole potential is not accounted for at $r>r_c$. We have shown that this simpler approach captures the correct behavior of the total ionization yields with molecular orientation and produces above threshold ionization spectra and photoelectron momentum distributions in agreement with the predictions from the full multielectron polarization potential. For atoms and molecules probed by strong laser fields, turning the external field off within the atomic/molecular radius solves a practical problem in single-active-electron TDSE calculations: It minimizes the dipole coupling of the active electron's orbital to the lower bound states of the potential, a coupling which is unphysical, because these orbitals are already occupied, but an effect that is not necessarily captured by the single-active-electron potential. Thereby, turning off the external field at $r<r_c$ increases the accuracy of the TDSE method within the single-active-electron approximation, and the approach is physical motivated by the polarization of the inner charge cloud counteracting the effect of the external field. We demonstrated the effectiveness of this method for  CO and reproduced the experimental distribution of total ionization yields. By applying the current approach to atomic systems such as Ar, there would be no need to include a hardcore boundary in the core region.

The present results have implications for future strong-field studies. One perspective is that the switching off procedure is easily implemented and therefore  single-active-electron potentials can readily be improved by suppressing unphysical coupling to occupied lower-lying bound states. A related perspective is that at least some typical observables of  strong-field ionization seem not to be very sensitive probes of the precise nature of the interactions involved. As an example, we have shown here that the multielectron polarization effect can be accurately modelled at the level of total ionization, by just knowing the static polarizability of the cation and the associated cut-off radius, $r_c$. For systems with larger polarizability this may change as the results could also be sensitive to the long-range part of the induced dipole potential (see, e.g., Refs.~\cite{PhysRevLett.106.073001,Dimitrovski_2015}). Similarly, these remarks are relevant for future investigations of holographic patterns in PMDs of small polar molecules aiming at extracting effects of Coulomb and electron-electron interactions, see for examples Refs.~\cite{PhysRevLett.108.263003,PhysRevLett.116.163004}: the holographic patterns imprinted in the PMDs of the CO molecule are not sensitive to the induced dipole term caused by the response of the CO$^+$ cation to the external field.

\begin{acknowledgments}
This work was supported by the Villum Kann Rasmussen (VKR) Center of Excellence, QUSCOPE - Quantum Scale Optical Processes. The numerical results presented in this work were obtained at the Centre for Scientific Computing (CSCAA), Aarhus. 
\end{acknowledgments}

\section{bibliography}

\end{document}